%% file: main.tex
\documentclass[
superscriptaddress,
amsmath,
amssymb,
pra,
twocolumn,
aps,
floatfix,
]{revtex4-1}

\input{packages.tex}

\begin{document}

\input{abstract.tex}

\title{Hybrid Quantum-Classical Eigensolver Without Variation or Parametric Gates}
\author{Pejman Jouzdani}
\email[Email:]{ jouzdanip@fusion.gat.com}
\thanks{corresponding author}
\affiliation{General Atomics, San Diego, CA}
\author{Stefan Bringuier}
\affiliation{General Atomics, San Diego, CA}
\date{\today}


\maketitle

\input{introduction.tex}

\input{background.tex}
\input{method.tex}

\input{numerical.tex}
\input{hardware.tex}
\input{discussion_summary.tex}
\input{acknowledgements.tex}

\bibliographystyle{apsrev}
\bibliography{references}
\pagebreak

\end{document}

%% file: packages.tex
\usepackage{graphicx}
\usepackage{epstopdf}
\usepackage{dcolumn}%
\usepackage{bm}%
\usepackage{chngcntr} 
\usepackage{hyperref}%
\usepackage{bbold}
\usepackage{physics}
\usepackage[justification=raggedright,singlelinecheck=false]{caption}
\usepackage{subcaption}
\usepackage{physics}

%% file: abstract.tex
\begin{abstract}
The use of near-term quantum devices that lack quantum error correction, for addressing quantum chemistry and physics problems, requires hybrid quantum-classical algorithms and techniques. Here we present a process for obtaining the eigenenergy spectrum of electronic quantum systems. This is achieved by projecting the Hamiltonian of a quantum system onto a limited effective Hilbert space specified by a set of computational bases. From this  projection an effective Hamiltonian is obtained. Furthermore, a process for preparing short depth quantum circuits to measure the corresponding diagonal and off-diagonal terms of the effective Hamiltonian is given, whereby quantum entanglement and ancilla qubits are used. The effective Hamiltonian is then diagonalized on a classical computer using numerical algorithms to obtain the eigenvalues. The use case of this approach is demonstrated for ground sate and excited states of BeH$_2$ and LiH molecules, and the density of states, which agrees well with exact solutions. Additionally, hardware demonstration is presented using IBM quantum devices for H$_2$ molecule. 
\end{abstract}

%% file: introduction.tex
\section{Introduction}
\label{sec:intro}


Quantum computers offer the ability to address problems in quantum many-body chemistry and physics by quantum simulation or in a hybrid quantum-classical approach. The latter method is considered the most promising approach for noisy-intermediate scale quantum (NISQ) devices \cite{Preskill2018}. The prospect and benefits of quantum algorithms, along with suitable hardware, is in overcoming the complexity of the wave-function of a quantum system as it scales exponentially with system size \cite{McArdle2020}. Therefore developing techniques and algorithms for NISQ era devices that may prove to have some computational advantage themselves, or establish a path towards ideas and foundations that provide advantage for future error-corrected quantum devices, is a worthwhile pursuit.

The leading algorithms intended to be executed on NISQ devices, which aim to determine solutions to an electronic Hamiltonian, are variational in nature \cite{Yuan2019}. One specific algorithm is the variational quantum eigensolver (VQE), which has been tremendously successful in addressing chemistry and physics problems on quantum hardware and NISQ devices \cite{Peruzzo2014,McClean2016,Kandala2017,Barkoutsos2018,Jones2019,Nakanishi2019,Parrish2019,Higgott2019,Jouzdani2019}. However, the restriction or challenge that exist with VQE is the need for prior insight with regard to selecting the trial quantum state, i.e., ansatz circuit. Furthermore, the classical optimization of the ansatz parameters may be a poorly converging problem \cite{McClean2018,Parrish2019b} and therefore limiting the applicability of VQE for obtaining results accurate enough for chemical or physical interpretation. Finally, the realization of ansatz circuits that are motivated by domain knowledge, for example the unitary coupled cluster ansatz for chemistry problems \cite{Romero2017}, may not be directly applicable on NISQ hardware and therefore requires clever modification to obtain hardware efficient ans\"atze \cite{ Kandala2017,Barkoutsos2018,Herasymenko2019}.

In this work, we present a pragmatic hybrid quantum-classical approach for calculating the eigenenergy spectrum of a quantum system within an effective model. Firstly, an effective Hamiltonian is obtained through measurement of  short-depth quantum circuits. The effective Hamiltonian is essentially the projection of the quantum system Hamiltonian onto a limited set of computational bases. The basis set is prepared with the intent of ensuring the dimensions of the corresponding matrix does not grow exponentially with the system size. In order to evaluate the matrix elements of the effective Hamiltonian,  suitable non-parametric quantum circuits are specified. The quantum circuits are designed, executed, and measured. From the result of the measurements, the diagonal and off-diagonal terms of the effective Hamiltonian matrix are obtained. On the classical side, the effective Hamiltonian matrix, with suitable dimensions, is diagonalized numerically using a classical computer.

The paper is organized as follows: A short background is presented in Section \ref{sec:background}. In Section \ref{sec:method} the steps taken in our hybrid quantum-classical approach are explained in detail.
In Section \ref{sec:numerical} we demonstrate the application of this hybrid approach on simple chemical molecules BeH$_2$ and LiH. In Section \ref{sec:hardware} the approach is demonstrated on the IBMQ 5-qubit Valencia quantum processor \cite{IBMQ-Valencia} for H$_2$ molecule. Finally, we discuss the integration of VQE and the proposed approach in Section \ref{sec:discussion_summary}.\par


%% file: background.tex
\section{Background}
\label{sec:background}

Consider a quantum many-body  system of electrons with the  second quantized Hamiltonian:
\begin{eqnarray}
\hat{H}  =  \sum_{ij} \kappa_{ij} a^{\dagger}_i a_j + \sum_{ijkl} v_{ijkl} a^{\dagger}_i a^{\dagger}_j a_k a_l.
\label{eq:subsec:hamiltonian:generic}
\end{eqnarray}
$a_i^\dagger$ and $a_i$ are the creation and annihilation operators, respectively. The anticommutator for the creation and annihilation are given by: $a_i a^\dagger_j  + a^\dagger_j a_i= \delta_{ij} $ and $
a_i a_j  + a_j a_i= a^\dagger_i a^\dagger_j  + a^\dagger_j a^\dagger_i = 0 $. These rules enforce the non-abelian group statistics for fermions, that is, under exchange of two fermions the wave-function yields a minus sign. \par

The indices in Eq.~(\ref{eq:subsec:hamiltonian:generic}) refer to single-electron states. The coefficients $\kappa_{ij}$ and $v_{ijkl}$ are the matrix integrals
\begin{eqnarray}
\kappa_{ij} &=& \bra{i} \hat{K}_1 \ket{j}
\label{eq:subsec:hamiltonian:matrix-elements-1}
\end{eqnarray}
and
\begin{eqnarray}
v_{ijkl} &=& \bra{ij} \hat{V}_{12} \ket{kl},
\label{eq:subsec:hamiltonian:matrix-elements-2}
\end{eqnarray}
where $\hat{K}_1$ and $\hat{V}_{12}$ operators correspond to one- and two-body interactions respectively.
Since $\hat{K}_1$ and $\hat{V}_{12}$ can depend on other parameters, such as the distance between nuclei, the Hamiltonian in  Eq.~(\ref{eq:subsec:hamiltonian:generic}) represents a class of problems. However, this class of problems has the common property that the number of fermions is a conserved value. Strictly speaking, the terms in the Hamiltonian act on fixed-particle-number Hilbert spaces, $\mathcal{H}_{N_F}$, that have the correct fermionic antisymmetry, with $N_F$ denoting the number of electrons. In this paper, we consider this class of electronic systems where the Hamiltonian is assumed to be in the form of Eq.~(\ref{eq:subsec:hamiltonian:generic}). The coefficients expressed in Eqs.~(\ref{eq:subsec:hamiltonian:matrix-elements-1}--\ref{eq:subsec:hamiltonian:matrix-elements-2}) can be obtained using software packages developed for quantum chemistry calculations that perform efficient numerical integration \cite{Reine2012}. \par

\subsection{Mapping to qubits \& computational basis}
\label{subsec:mapping}

The Hamiltonian as written in Eq.~(\ref{eq:subsec:hamiltonian:generic}) can be expressed in the form of qubit operations (i.e., Pauli matrices). This requires a transformation that preserves the anticommutation of the annihilation and creation operators. One transformation that satisfies the criteria, and is based on the physics of spin-lattice models, is the Jordan-Wigner (JW) transformation \cite{Fradkin1989}. The JW-transformed Hamiltonian takes the form

\begin{eqnarray}
\hat{H}  =  \sum_{s} \lambda_s \hat{h}_s,
\label{eq:subsec:mapping:JW-H}
\end{eqnarray}
in which $\lambda_s$'s are scalar and a \emph{Pauli string} operator $\hat{h}_s$ is defined as
\begin{eqnarray}
\hat{h}_{s}  =  \hat{O}^{s}_1 \otimes \cdots \otimes \hat{O}^{s}_N.
\label{eq:subsec:mapping:pauliop}
\end{eqnarray}
$\hat{O}^{s}_i \in \{\hat{I}, \hat{X}, \hat{Y}, \hat{Z} \}$ acts on the $i$-th qubit, $\{ \hat{X}, \hat{Y}, \hat{Z} \}$ are the three Pauli matrices \cite{Nielsen2002} and  $\hat{I}$ is the identity matrix with the number of qubits denoted as $N$.\par

If the number of $\hat{I}$ operators in the tensor product of $\hat{h}_s$ is $N-k$, we call $\hat{h}_s$ a $k$-local Pauli string operator. Upon the JW transformation of the Hamiltonian, a Fock basis of the second quantization representation is in one-to-one correspondence with a computational basis of the qubits \cite{Steudtner2019}. In other words, a computational basis of 
\begin{eqnarray}
\ket{\bf n} = \ket{n_0, n_1, \dots, n_N},
\label{eq-fock}
\end{eqnarray}
with $N$  qubits, where $n_i\in\{0,1\}$, is equivalent to a an antisymmetric Fock basis.

Within the finite, but exponentially large, Hilbert space spanned by  $2^N$ computational basis set, an effective matrix representation of the Hamiltonian may be possible, specifically, if one can efficiently evaluate the matrix elements $\bra{\bf n^\prime} \hat{H} \ket{\bf n}$ for an arbitrary computational basis $\ket{\bf n}$ and $\ket{\bf n^\prime}$. Furthermore, assuming that the dimensions of the resulting effective matrix are relatively small, the matrix can be diagonalized on a classical computer, where its eigenvalues approximate the spectrum of the original Hamiltonian. \par

In this paper, we show how to evaluate a matrix element $\bra{\bf n^\prime} \hat{H} \ket{\bf n}$ for arbitrary computational basis $\ket{\bf n}$ and $\ket{\bf n^\prime}$, using a quantum circuit that has a circuit depth $\mathcal{O}(N)$. We do so by using ancilla  qubits, and thus $N+1$ physical resources are needed. 
In addition, we  discuss how to choose an effective subspace for a given electronic Hamiltonian, with a dimension $N_s$, based on physical motivations (see Section~\ref{subsec:fockspace}).
The condition $N_s << 2^N$ makes it possible to diagonalize the Hamiltonian on a classical computer. In Section~\ref{sec:numerical}, we numerically demonstrate this method for simple quantum chemistry systems, focusing on ground state energy and the density of state calculations of the low-energy spectrum. Furthermore, in Section~\ref{sec:hardware} we use the IBMQ 5-qubit Valencia device to measure the terms for H$_2$ in the complete computational basis of 4 qubits. \par

%% file: method.tex
\section{Constructing  An Effective Matrix Representation for a Hamiltonian by Qubit Measurement}
\label{sec:method}
\subsection{Effective Hamiltonian and circuit representation}
\label{subsec:idea}

We first consider a  Hamiltonian $\hat{H}$ that is expressed in terms of Pauli strings as in Eq.~(\ref{eq:subsec:mapping:JW-H}). Additionally, a subspace $\mathcal{S} = \{ \ket{\bf n} \}$, with $N_s$ corresponding computational bases, is considered such that $N_s<<2^N$. Let us define  the \emph{effective Hamiltonian} matrix as the projection of $\hat{H}$ onto this subspace; that is 
\begin{eqnarray}
\hat{H}_{eff} = \sum_{\bf n, n^\prime \in \mathcal{S}} 
\bra{\bf n} 
\hat{H} \ket{\bf n^\prime} \,\, \ket{\bf n}
\bra{\bf n^\prime}.
\label{eq:subsec:idea:effective-H}
\end{eqnarray}
The next step is to define  a simple quantum circuit that utilizes one ancillary qubit to measure 
$\bra{\bf n} \hat{H} \ket{\bf n^\prime}$ matrix element. \par

The dimension of $\hat{H}_{eff}$ depends on the choice of the subspace in $\mathcal{S}$. The choice of the subspace, intuitively,  depends on the physics of the problem. However, the focus of this paper is towards quantum chemistry problems, which are a primary application for NISQ devices. For this class of Hamiltonians there is a systematic way to select the appropriate subspace. This is discussed in Section~\ref{subsec:fockspace}. \par

The evaluation of diagonal terms in $\hat{H}_{eff}$, e.g.,  $\bra{\bf n} \hat{H} \ket{\bf n}$, is trivially performed by preparing $N$ qubits as a bit string of $\ket{\bf n} \equiv (n_0, \dots, n_N)$ and measuring $\hat{H}$. The measurement of the total Hamiltonian is obtained by measuring every individual Pauli string, $\hat{h}_s$, in Eq.~(\ref{eq:subsec:mapping:pauliop}). The diagonal terms are then given by
\begin{eqnarray}
\bra{\bf n} \hat{H} \ket{\bf n} = \sum_{s} \lambda_s\bra{\bf n} \hat{h}_s \ket{\bf n}.
\label{eq:subsec:idea:effecitve-H-diag}
\end{eqnarray}

\begin{figure}
\input{Figure-CoreCircuit-Standalone.tex}
\caption{Quantum circuits  for measuring (a) the \emph{real} 
and 
(b) 
\emph{imaginary} parts of an off-diagonal element $\bra{\bf n^\prime} \hat{H} \ket{\bf n}$. 
(c) A controlled-$n$ gate represents a set of \textsc{CNOT} gates that prepares the qubits in the state $\ket{n}$.  An example of controlled-$n$ for $N=3$ and $\ket{n} = \ket{011}$ is shown.}
\label{fig:subsec:idea:core-circ}
\end{figure}
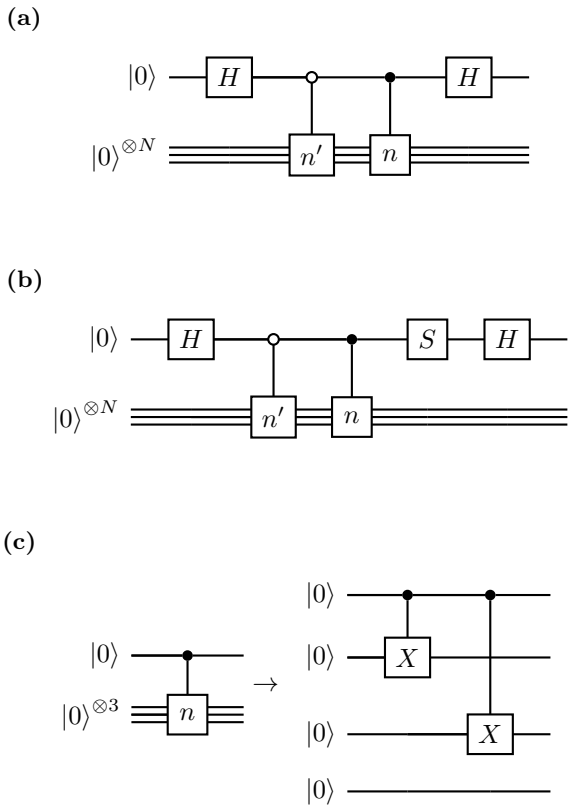

The off-diagonal matrix elements, e.g., $\bra{\bf n^\prime}\hat{H}\ket{\bf n}$, which are generally complex numbers, can be evaluated using a single ancillary qubit. This can be done by considering the quantum circuits as shown in Fig.~\ref{fig:subsec:idea:core-circ}. The two circuits shown in Fig.~\ref{fig:subsec:idea:core-circ}(a) and (b) are used to calculate the real and imaginary parts of the matrix element, respectively. In both circuits the $N+1$ qubits are initially prepared in $\ket{\psi_{int}} = \ket{0} \otimes \ket{0}^{\otimes N}$ state. The qubits are assumed to be enumerated linearly from $1$ to $N+1$, where the $N+1$-th qubit is the control qubit. \par

After applying the first Hadamard gate on the ancillary qubit, from left to right as shown in the circuits in Fig.~\ref{fig:subsec:idea:core-circ}(a--b), the quantum state of all the qubits is
\begin{eqnarray}
\ket{\psi} = \frac{1}{\sqrt{2}}\left[\ket{0} \ket{0}^{\otimes N} + \ket{1} \ket{0}^{\otimes N}\right],
\label{eq:subsec:idea:initial-quantumstate}
\end{eqnarray}
which, after a sequence of controlled-$X$ gates, becomes entangled as
\begin{eqnarray}
\ket{\psi} = \frac{1}{\sqrt{2}}
\left[\ket{0} \ket{\bf n^\prime} + \ket{1} \ket{\bf n}\right].
\label{eq:subsec:idea:final-quantumstate}
\end{eqnarray}
Here, control gates (controlled-$n$ and controlled-$n^\prime$) flip $N_F$ qubits (corresponding to the $N_F$ occupied electronic states) and prepares the target qubits in the computational basis $\ket{\bf n}$ ($\ket{\bf n^\prime}$), conditioned on the state of the control qubit is $\ket{1}$ ($\ket{0}$). \par

An example of a controlled-$n$ gate that prepares target qubits in $\ket{011}$ state is shown in Fig.~\ref{fig:subsec:idea:core-circ}(c). In practice, this part of the circuit requires two-qubit gates (e.g., \textsc{CNOT}) and perhaps the need for full connectivity of qubits in order to operate on any two  qubits. Full connectivity of qubits could potentially be realized with ion-trapped devices \cite{Wright2019}. See Section~\ref{sec:discussion_summary} for further discussion.\par

Depending on whether the \emph{real part} or the \emph{imaginary part} is calculated,  the last gates acting on the control qubit changes. With regard to the \emph{real part}, after applying the last Hadamard gate on the control qubit in Fig.~\ref{fig:subsec:idea:core-circ}(a), the quantum state is
\begin{eqnarray}
\ket{\psi} = \frac{1}{2}
\ket{0} \left[ \ket{\bf n^\prime}+\ket{\bf n}\right] + 
\frac{1}{2}\ket{1} \left[ \ket{\bf n^\prime}-\ket{\bf n}\right].
\end{eqnarray}
Using this prepared quantum state, one can measure
\begin{eqnarray}
\hat{M}_{0} &=& \ket{0} \bra{0} \otimes \hat{H},
\label{eq:subsec:idea:what-is-measured}
\end{eqnarray}
at the end of the circuit and have
\begin{eqnarray}
m_{0} &=& \bra{\psi} \left[\ket{0} \bra{0} \otimes \hat{H} \right] \ket{\psi} \nonumber \\
&=& \frac{1}{2^2} \left(\bra{\bf n} \hat{H} \ket{\bf n} + 
\bra{\bf n^\prime} \hat{H} \ket{\bf n^\prime} +
2 {\bf Re} \left[\bra{\bf n} \hat{H} \ket{\bf n^\prime}) \right] \right). \nonumber \\
\label{eq:subsec:idea:measurement-theory}
\end{eqnarray}
Thus, after substituting for the  diagonal elements $\bra{\bf n} \hat{H} \ket{\bf n} $  and $\bra{\bf n^\prime} \hat{H} \ket{\bf n^\prime}$, using Eq.~(\ref{eq:subsec:idea:effecitve-H-diag}), one obtains the real part of the off-diagonal matrix element $ \bra{\bf n} \hat{H} \ket{\bf n^\prime}$.
The value $m_0$ in Eq.~(\ref{eq:subsec:idea:measurement-theory}) is measured on a quantum device using the identity $\ket{0} \bra{0}=\frac{1}{2}(\hat{I}+\hat{Z})$, by 
\begin{eqnarray}
m_{0} &=& 
\frac{1}{2} \bra{\psi} \hat{I} \otimes \hat{H} \ket{\psi}
+\frac{1}{2} \bra{\psi} \hat{Z} \otimes \hat{H} \ket{\psi}
\nonumber \\
&=& 
\sum_{s}  
\frac{\lambda_s}{2} \bra{\psi} \hat{I} \otimes \hat{h}_s \ket{\psi}
+\frac{\lambda_s}{2} \bra{\psi} \hat{Z} \otimes \hat{h}_s \ket{\psi}.
\label{eq:subsec:idea:measurement-device}
\end{eqnarray}
\par

The \emph{imaginary part} can be obtained in a similar fashion as done for the \emph{real part}, but with a slight modification to the circuit as shown in Fig.~\ref{fig:subsec:idea:core-circ}(b). The key addition is a phase-gate, $S$,  before the execution of the last Hadamard gate on the control qubit. This yields the quantum state
\begin{eqnarray}
\ket{\psi} = \frac{1}{2}
\ket{0} \left[ \ket{\bf n^\prime}+i\ket{\bf n}\right] + 
\frac{1}{2}\ket{1} \left[ \ket{\bf n^\prime}-i\ket{\bf n}\right]
\end{eqnarray}
that now includes a phase factor $i$ before $\ket{n}$. After the last Hadamard gate in Fig.~\ref{fig:subsec:idea:core-circ}(b), and following the same steps, Eqs.~(\ref{eq:subsec:idea:what-is-measured})-(\ref{eq:subsec:idea:measurement-device}), the imaginary part ${\bf Im} \left[\bra{\bf n} \hat{H} \ket{\bf n^\prime}) \right]$ is obtained. \par

Our approach differs from the typical hybrid quantum-classical paradigm used in ground state chemistry electronic structure calculations in that the  quantum hardware is used as a coprocessor to measure these matrix elements. Therefore, no parameterized ansatz  or variational optimization is required. In this approach, the depth of the quantum circuit is significantly reduced, however, our method is based on the assumption that the dimensions of $\hat{H}_{eff}$ in Eq.~(\ref{eq:subsec:idea:effective-H}) are reasonable enough such that it can be diagonalized using  classical numerical algorithms.\par

\subsection{Implementing Measurements}
\label{subsec:measurement}
As shown in Eq.~(\ref{eq:subsec:idea:measurement-device}), the expectation value of the Hamiltonian becomes the weighted sum of expectations for the set of Pauli string operators with respect to the output quantum state of the circuit. Since these operators are not in general commuting, one needs to setup and run a number of different quantum circuits. This number can be  up-to all Pauli string operators in the Hamiltonian. Each circuit is then executed many times and every time the qubits are measured, the results are realized in different computational bases. The sampled realization provides a probability distribution and is used to estimate the expectation value of the Pauli string.\par

Generally, measurements are done either directly or indirectly. In the case of direct measurements, single-qubit rotations are applied to a subset of qubits at the end of the circuit. This subset is identified based on the locations in the tensor-product of the operator $\hat{h}_s$ that are not identity $\hat{I}$. This set of rotations essentially changes the computational bases in which the given operator $\hat{h}_s$ is diagonal.
The direct measurement is commonly used in the experimental demonstration of quantum hardware and VQE \cite{Kandala2017}. \par

The indirect measurement approach \cite{Knill2007} requires a series of controlled gates that are applied to the $N$ target qubits, using one ancillary control qubit. The indirect measurement method is used in iterative quantum phase estimation algorithms \cite{Miroslav2007}. \par 

Although the two types of measurement approaches are theoretically equivalent \cite{Nielsen2002}, experimentally, there are differences. The benefits and drawbacks of direct and indirect measurements are discussed for example in \cite{Mitarai2019}. The main difference is the number of times the circuit is to be executed to achieve a desired precision $\epsilon$, which is $\mathcal{O}(\frac{1}{\epsilon^2})$ and $\mathcal{O}(\frac{1}{\epsilon})$ for direct and indirect measurements respectively. The implementation of a general control-$U$ gate in the indirect measurement is a challenging task \cite{Knill2007}. However, in regards to our purposed approach, where the $U$ operator is a single Pauli string, the indirect measurement implementation is straightforward.\par

\begin{figure}[ht]
  \includegraphics{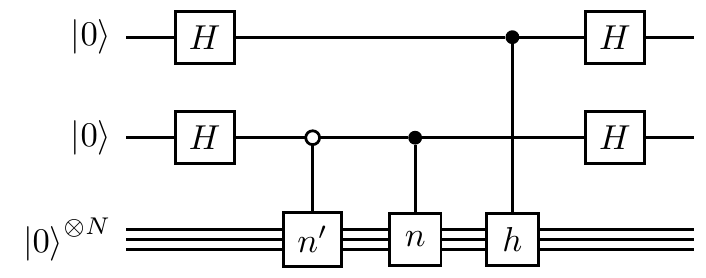}
  \caption{Quantum circuit  for indirect measurment of the \emph{real} and \emph{imaginary} parts of an arbitrary element $\bra{\bf n^\prime} \hat{h}_s \ket{\bf n}$, for a given Pauli string operator $\hat{h}_s$. Only controlled one-qubit gates are used. The first control qubit from the top is used for the measurement of the operator $\hat{h}_s$. The second control is to prepare the quantum state of the target qubits.}
\label{fig:subsec:measurement:core-circ}
\end{figure}

The indirect measurement can be adapted as shown in quantum circuit in Fig.~\ref{fig:subsec:measurement:core-circ}. An additional control qubit is added to this circuit. The  Pauli operators are applied on the $k$ locations of $\hat{h}_s$ when the control is in state $\ket{1}$. It is straightforward to follow the quantum state of the qubits throughout the circuit. The final state  reads as:
\begin{eqnarray}
\ket{\psi} = \frac{1}{2}
\ket{00}
\left[
\ket{\bf n^\prime}+
\ket{\bf n}+
\hat{h}\ket{\bf n^\prime}+
\hat{h} \ket{\bf n}
\right] + \dots \nonumber .\\
\label{eq:subsec:measurement:state-indirect}
\end{eqnarray}
Upon measuring $\hat{M} = \ket{0}\bra{0} \otimes \ket{0}\bra{0} \otimes 
\hat{I}$, at the end of this circuit,  using Eq.~(\ref{eq:subsec:measurement:state-indirect}), and steps discussed in sec.~\ref{subsec:idea}, the real part of the off-diagonal matrix element $\bra{\bf n^\prime} \hat{h}_s \ket{\bf n}$ is obtained. The imaginary part is determined  under same steps, while an additional phase gate is applied to the last ancillary qubit, similar to the situation in Fig.~\ref{fig:subsec:idea:core-circ}(b).\par

In terms of Pauli strings, the value of $\bra{\psi} \hat{M} \ket{\psi} $ is obtained from the expectations
$ \bra{\psi} \hat{Z}_{N+2}  \ket{\psi}$, 
$\bra{\psi} \hat{Z}_{N+1}   \ket{\psi}$, 
and  
$\bra{\psi} \langle\hat{Z}_{N+2}\otimes \hat{Z}_{N+1}  \ket{\psi}$, similar to Eq. (\ref{eq:subsec:idea:measurement-device}).

Finally, using the state preparation and measurements outlined through Sections~\ref{subsec:idea}--\ref{subsec:measurement}, all the $N_s \times N_s$ matrix elements of the effective Hamiltonian in Eq.~(\ref{eq:subsec:idea:effective-H}) can be evaluated, by repeating the execution of the quantum circuits as discussed in Figs.~\ref{fig:subsec:idea:core-circ}-\ref{fig:subsec:measurement:core-circ}, for all the possible combinations of a chosen set of $\{\ket{\bf n}\}$ bases. 

Note that the approach for evaluating the \emph{real} and \emph{imaginary} parts of a matrix element is similar to the interference method introduced in ref.~\cite{Parrish2019}, with the difference being that our approach uses an ancillary qubit in order to realize the interference. \par

\subsection{Preparing the computational basis}
\label{subsec:fockspace}
The computational basis set  $\mathcal{S}= \{ \ket{\bf n} \}$ with  $N_s<<2^N$, needs to be specified in practice. These bases serve as the row and columns of $\hat{H}_{eff}$  in Eq.~(\ref{eq:subsec:idea:effective-H}). The process and motivation for how to choose this set should be based on the underlying nature and physics of the problem.\par

In theory, one established approach to approximate the ground states of quantum many-body systems is  mean-field theory \cite{Jensen2017}. In this approach, the true ground state is constructed by perturbing a reference mean-field quantum state. The quantum chemistry field has established theories and techniques for treating such problems. One particularly successful theory and numerical method is coupled cluster (CC), typically referred to as the gold-standard in computational quantum chemistry \cite{Bartlett2007}. In CC one assumes a wave-function ansatz
\begin{eqnarray}
\ket{ \psi} &=& e^{\hat{T}} \ket{\bf 0},
\label{eq:subsec:fockspace:cc-ansatz}
\end{eqnarray}
for the ground state. Here $\ket{\bf 0}$ is a reference quantum state (e.g., Hartree-Fock) and is considered to be anti-symmetric under exchange of two fermions. The operator $\hat{T} = \hat{T}_1 + \hat{T}_2 + \cdots$ is a sum over different possible excitation operators with respect to the reference state. Typically, the set of excitation operators in $\hat{T}$ includes single and double terms (i.e., CCSD), which enables a series representation of the Taylor expansion of $e^{\hat{T}}$, but high-order terms can also be added. The coefficients for the excitation operators inside $\hat{T}$ are determined by variational methods; that is, by minimizing the expectation of the Hamiltonian with respect to the ansatz \cite{Bartlett2007}. \par 

Since the exponential operator in the CC ansatz, Eq. (\ref{eq:subsec:fockspace:cc-ansatz}), is non-unitary, it cannot be directly implemented on gate-based quantum computers, where gates correspond to unitary operators. Thus, a unitary version of the CC ansatz has been introduced and is known as Unitary Coupled Cluster (UCC) \cite{Romero2017,Wecker2015}. Ideally, the implementation of UCC ansatz should be constructed such that the number of gates is minimized so that the circuit depth does not exhaust the current coherence times of NISQ devices. As a result of the this concern, hardware efficient anstaz have been proposed \cite{Kandala2017} as a substitute. \par

In the numerical demonstration of this work (see Section~\ref{sec:numerical}), we consider a simplified ansatz for the ground state as:
\begin{eqnarray}
\ket{ \psi} &=& 
c_{0} \ket{\bf 0} 
\nonumber \\
&+& \sum_{i \nu} c_{i\nu} a_i^\dagger a_\nu \ket{\bf{0}} \nonumber \\
&+& 
\sum_{ij \nu \beta} c_{ij \nu \beta} a_i^\dagger a^\dagger_j a_\nu a_\beta \ket{\bf{0}} 
\nonumber \\
&+& \cdots,
\label{eq:subsec:fockspace:ucc-ansatz}
\end{eqnarray}
where $i,j,\dots$  refer to the unoccupied levels, and $\nu, \beta, \dots$, refer to occupied levels with respect to single (\textit{S}), double (\textit{D}), and higher order excitation operators. The ansatz in Eq.~(\ref{eq:subsec:fockspace:ucc-ansatz}) implies that the true many-body ground state is a superposition of the reference state $\ket{\bf 0}$ and all possible \textit{S}
$\{a_i^\dagger a_\nu \ket{\bf{0}}\}$, \textit{D} $\{a_i^\dagger a^\dagger_j a_\nu a_\beta \ket{\bf{0}}\}$, up-to   $\hat{T}_{n}\ket{\bf 0}$ excitations, where $n$ in $\hat{T}_{n}$ is finite and independent of $N$.\par

In particular, assuming that it is possible to truncate the series at some low-excitation level such as the \textit{D} or triples (\textit{T}), the number of eigenstates in the expansion of the above ansatz remains a polynomial function in $N$. \par

Taking the ansatz in Eq.~(\ref{eq:subsec:fockspace:ucc-ansatz}), we specify the set $\mathcal{S}$ in the following way. (1) Pick a  computational basis as the reference quantum state, that is  $\ket{\bf 0}  = \ket{\bf n_{int} }$. (2) Identify computational bases corresponding to a finite number of excitations such as \textit{S} and  \textit{D}. These are  $N_s<<2^N$  bases and  polynomial function in $N$.\par

In step (1), we identify the initial computational basis $\ket{\bf n_{int}}$ by minimizing the 
$\bra{\bf n_{int}} \hat{H}\ket{\bf n_{int}}$, in which, for example, a classical Monte Carlo process from spin lattice models \cite{Janke1996} can be used. This computational basis is essentially the qubits'  configuration that has the lowest energy expectation. Since this step is a classical one, it is performed effectively even for a  large number of qubits. 

In step (2), once the state $\ket{\bf n_{int}}$ is determined, one can rearrange the configuration of the qubits by swapping $1$'s and $0$'s within the state $\ket{\bf n_{int}}$. The swapping is done so that the configurations  corresponding to \textit{S}, \textit{D}, up-to $\hat{T}_n$ are fully realized. The energy expectation  corresponding to these configurations (the diagonal elements in the $H_{eff}$) are stored on a classical register. The final result is obtained among the set of configurations whereby $N_s$ of are the lowest energies; these are the configurations that are selected. The above steps are demonstrated numerically which is discussed in the next section (see Section~\ref{sec:numerical}). \par

%% file: Figure-CoreCircuit-Standalone.tex

\captionsetup[subfigure]{position=top, labelfont=bf,textfont=normalfont,justification=raggedright,format=plain}
\begin{subfigure}{0.45\textwidth}
   \caption{}
   \includegraphics{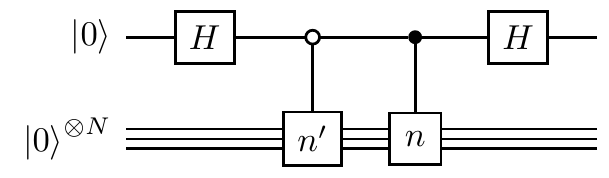}
   \label{subfig:sub1:fig:core-circ}
\end{subfigure}

\vspace{1cm}

\begin{subfigure}{0.45\textwidth}
  \caption{}
  \includegraphics{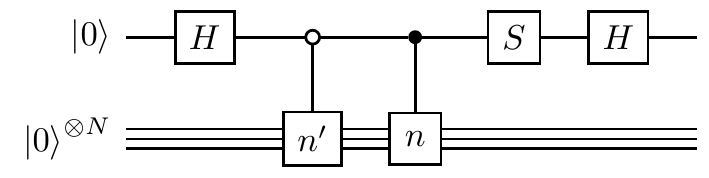}
  \label{subfig:sub2:fig:core-circ}
\end{subfigure}

\vspace{1 cm}

\begin{subfigure}{0.45\textwidth}
  \caption{}
    \includegraphics{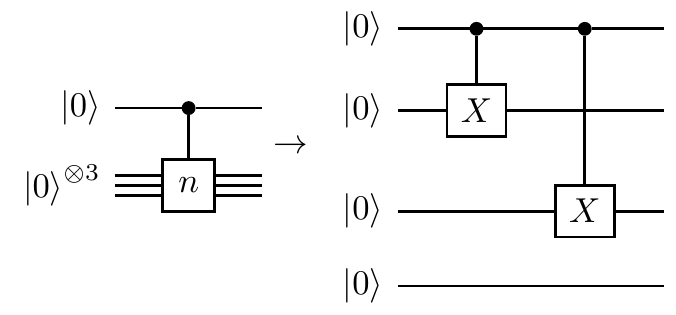}
  \label{subfig:sub3:fig:core-cirq}
\end{subfigure}

%% file: numerical.tex
\section{Numerical Demonstration: LiH and BeH$_2$}
\label{sec:numerical}

The application of the methodology discussed in Section~\ref{sec:method} is focused on BeH$_2$ and LiH molecules due to thier relatively small number of electrons and molecular orbital footprint. The number of electrons in BeH$_2$ and LiH is six and four, respectively. The single-electron molecular spin-orbitals in the second quantized Hamiltonian are constructed using the minimal atomic STO-3G basis set \cite{Kandala2017,McArdle2020}. For BeH$_2$ there are a total of $14$ spin-orbital, and thus corresponds to $14$ qubits; LiH contains $12$ spin orbitals and hence $12$ qubits. For the purpose of measuring the matrix elements of the effective Hamiltonian, Eq.~(\ref{eq:subsec:idea:effective-H}), one additional ancillary qubit is required, as illustrated in the quantum circuits shown in Fig.~\ref{fig:subsec:idea:core-circ}. Therefore, the total number of qubits for the simulation of BeH$_2$  and LiH is $15$ and $13$, respectively. \par 

To obtain the coefficients in the Hamiltonian of Eq.~(\ref{eq:subsec:hamiltonian:generic}) --- more specifically as defined in Eqs. (\ref{eq:subsec:hamiltonian:matrix-elements-1}) and (\ref{eq:subsec:hamiltonian:matrix-elements-2}) --- we make use of the Psi4 quantum chemistry package \cite{Psi4}. The second quantized Hamiltonian is further transfomed onto a set of Pauli strings and their corresponding weights, Eq.~(\ref{eq:subsec:hamiltonian:generic}), via JW transform using OpenFermion package \cite{OpenFermion}. \par

In order to construct the potential energy surface for each inter-nuclear distance, $R$, the steps indicated in the previous paragraph are repeated. The distance $R$ corresponds to the bond length between Be--H or Li--H in a given molecule; both LiH and BeH$_2$ are linear molecules. We note that these calculations are assuming the total Hamiltonian can be represented using the Born-Oppenheimer approximation, where  the dynamics of the core nuclei are neglected. This is standard practice in quantum chemistry calculations \cite{Jensen2017} and thus at every distance $R$, the nuclear-nuclear repulsion energy is treated classically and is added to the Hamiltonian as a constant. \par

\begin{figure}
\centering
\includegraphics[width=0.45\textwidth]{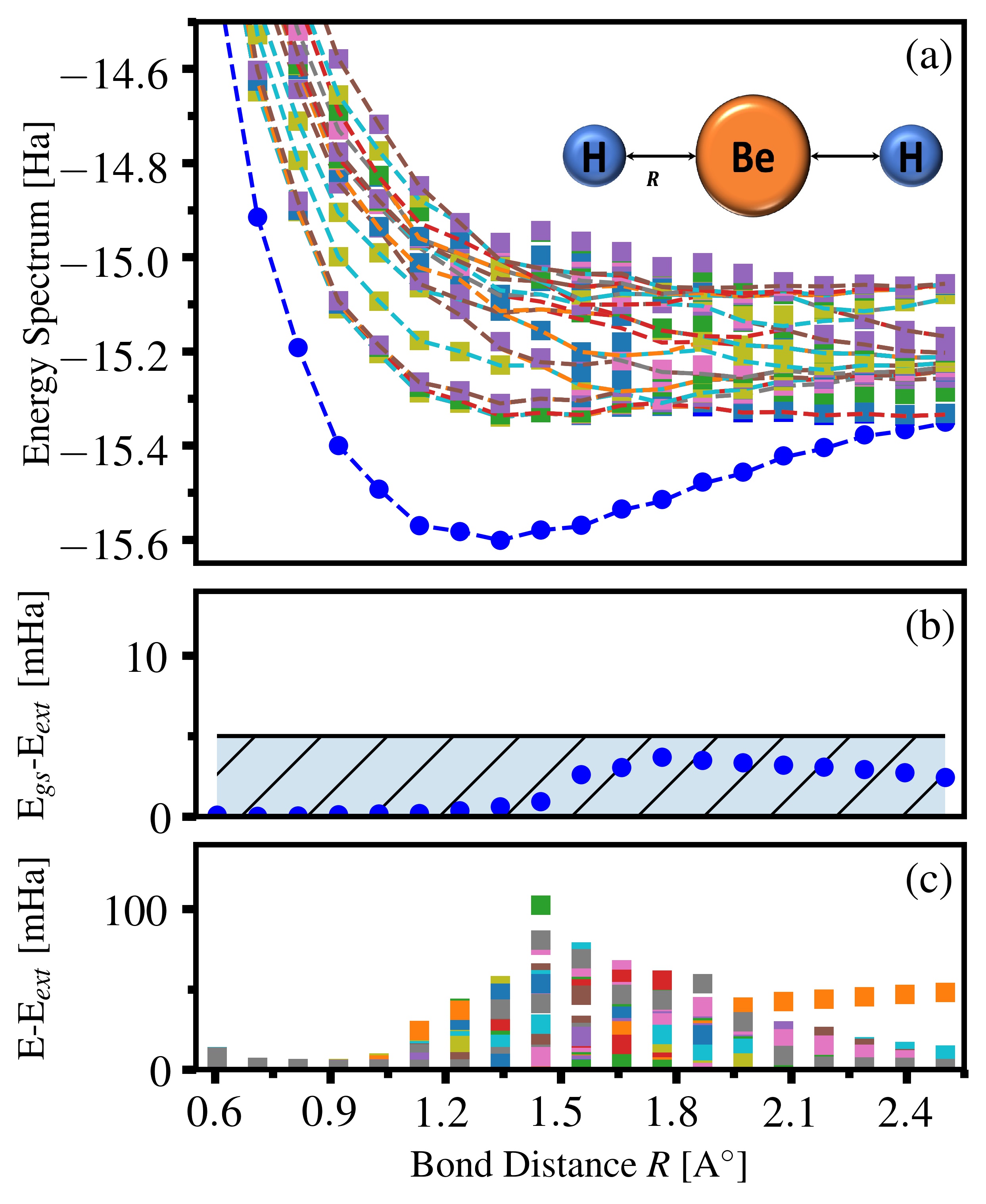}
\caption{(a) The calculated binding curve of BeH$_2$ for the ground state and several excited states demonstrating the application of the method described in this work. (b) The difference between the ground state energies obtained from diagonalization of
the exact and the effective Hamiltonian along with the chemical accuracy line, and (c) is energy difference for the obtained excited state spectra within the effective model.}
\label{fig:numerical:BeH2-Curve}
\end{figure}

For every distance $R$, a set of computational bases is chosen. This set includes the basis with the lowest energy expectation (the reference configuration) and computational bases that correspond to the low-order excitations (i.e.,  \textit{S}, \textit{D}, etc.) with respect to the reference configuration.  In the case of BeH$_2$, we construct the effective Hamiltonian matrix by keeping the \textit{S}, \textit{D}, and \textit{T} excitations, and thus the total number of computational bases for BeH$_2$ is $N_{s}=1588$. The low dimension of the  effective Hamiltonian matrix makes it possible to diagonalize the matrix on a classical computer using standard numerical techniques \cite{Scipy}. Thus, the lowest energies, including the ground state energy, are obtained at every $R$. The results for this process are shown in Fig.~\ref{fig:numerical:BeH2-Curve}.  \par

\begin{figure}
\centering
\includegraphics[width=0.45\textwidth]{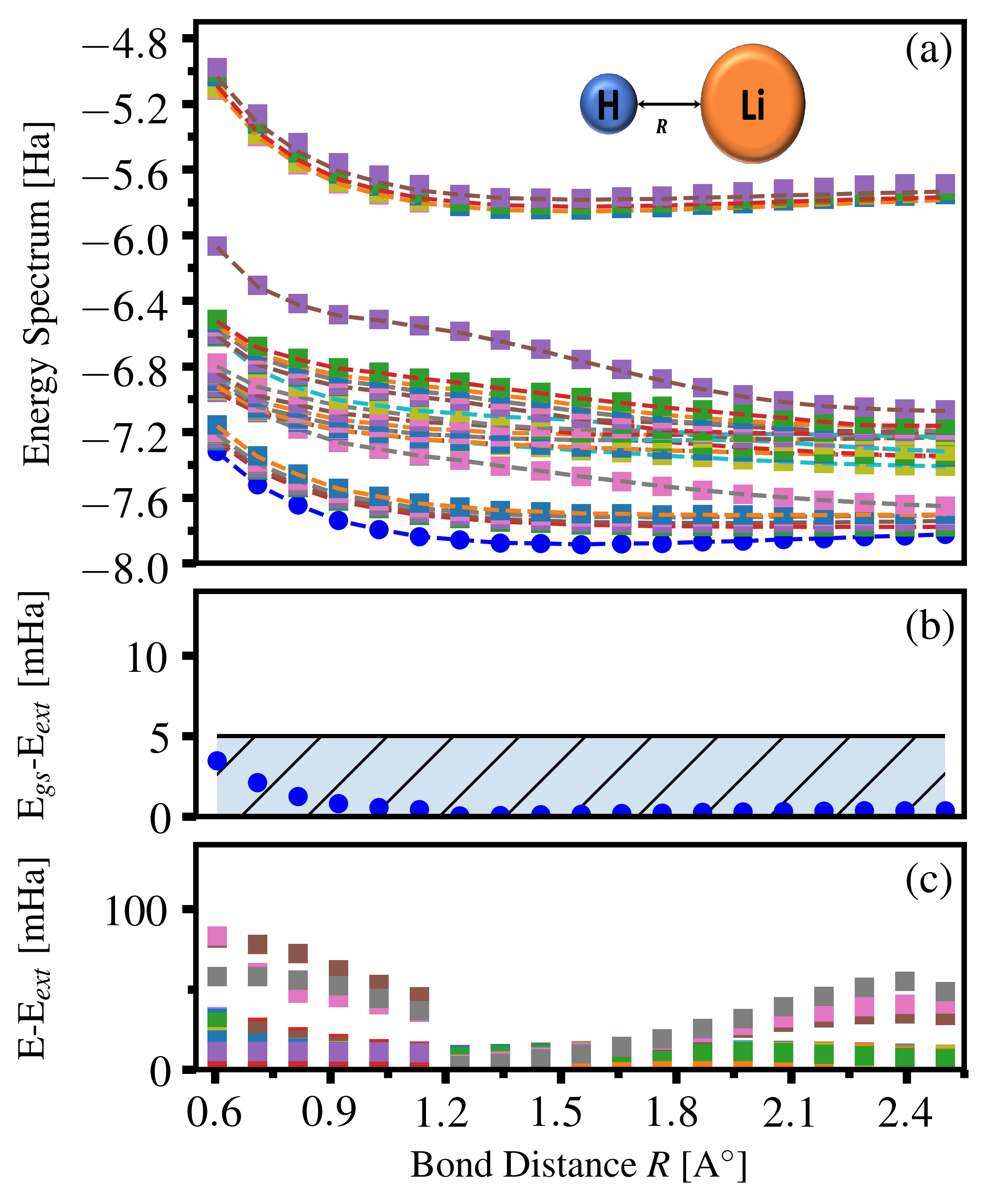}
\caption{ Similar result as shown in Fig.~\ref{fig:numerical:BeH2-Curve} but for LiH. (a) Shows ground state and excited states and (b-c) indicate energy differences compared to exact diagonalization.}
\label{fig:numerical:LiH-Curve}
\end{figure}

Fig.~\ref{fig:numerical:BeH2-Curve}(a) shows the ground state energy  as well as a few low-lying excited states for BeH$_2$ of the effective Hamiltonian. The exact energies are given by the dashed curves in Fig.~\ref{fig:numerical:BeH2-Curve}(a). In Fig. \ref{fig:numerical:BeH2-Curve}(b), the difference between the exact and obtained ground state energy is shown. An error within the hatched area indicates results that are within chemical accuracy. Chemical accuracy is typically identified as $\sim 5\times 10^{-3}$ Hatrees. Figure~\ref{fig:numerical:BeH2-Curve}(c) shows the same energy difference for all other excited-state energies. 
The same process is done for LiH where only \textit{S} and \textit{D} configurations are used. The total number of permutations of $4$ fermions ($495$) is reduced to $N_s = 200$, and the results are shown in Fig.~\ref{fig:numerical:LiH-Curve}. \par

The total number of configurations can be approximated by the highest excitation considered. For $n$-electron excitation the number of configurations is $\binom{N_F}{n}\binom{N-N_F}{n}< N^{2n}$. Thus, the total number of configurations is less than $1+ \cdots+N^{2n} = \mathcal{O}(N^{2n})
$ when $N \gg 1$.
Of course, this argument is valid as long as $n$ can be assumed to be finite and independent of the size of the system $N$ or the number of electrons $N_F$. Under these assumptions, the dimensions of the effective Hamiltonian, $H_{eff}$, remains polynomial if one replaces the minimal STO-3G with the extended atomic basis.\par

\subsection{Density of States}
\label{subsec:dos}
\begin{figure}[h]
\centering
\includegraphics[width=0.45\textwidth]{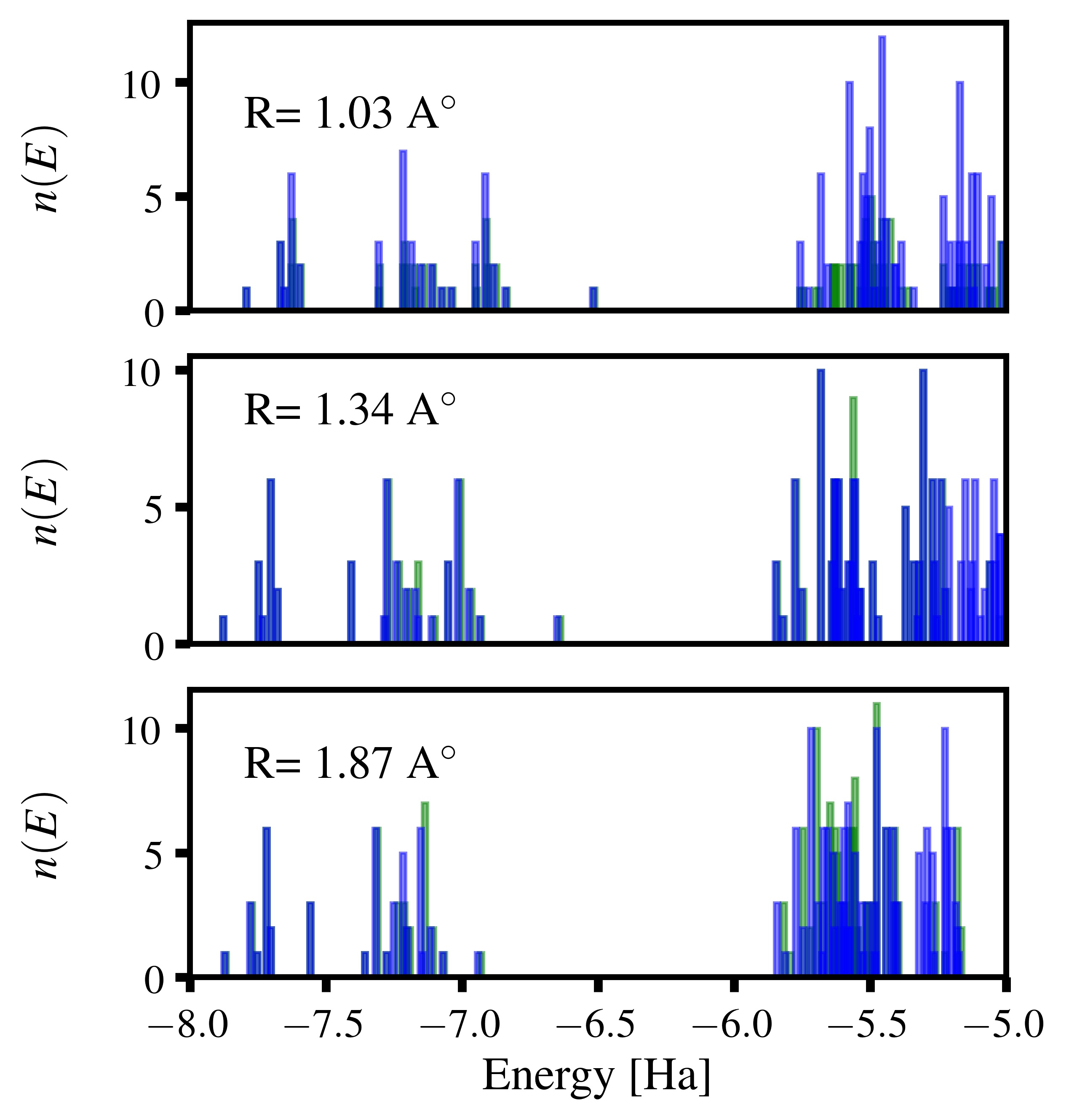}
\caption{The unnormalizaed density of states for LiH at different binding distances. The blue fill indicates the exact and the green fill indicates simulated}
\label{fig:numerical:LiH-DOS}
\end{figure}

Knowledge of density of states (DOS) can be important in analyzing the thermodynamic behaviour of a system at finite temperature and in the analysis of transition states  important in chemical reactions \cite{Truhlar1996}. One advantage of the method proposed in this paper is the insights provided into the low-energy spectrum of the quantum system, more precisely, the degeneracy of the energy levels. To illustrate this, in Fig.~\ref{fig:numerical:LiH-DOS}, the unnormalized DOS of the LiH obtained via the effective Hamiltonian is compared with the exact density, which shows qualitatively good agreement. \par

%% file: hardware.tex
\section{Hardware Demonstration: H$_2$}
\label{sec:hardware}
In this section, we demonstrate the feasibility of the hybrid quantum-classical approach discussed in this paper on a IBMQ hardware device. Specifically, we calculate the eigenenergy spectrum of H$_2$ molecule using  STO-3G minimal basis-set. 

Within STO-3G basis set, the total number of spin-orbitals for H$_2$ is four. In contrast to the numerical demonstration, where the computational basis was prepared using the UCC ansatz, here we employ the complete computational basis with conserved number of electrons for H$_2$ ($N_F=2$). The number of computational basis set with conserved electron number, is $N_s=6$, and the bases are:
\begin{eqnarray}
\{\ket{n_0, n_1, n_2, n_3}\} =& \{\ket{1100}, \ket{1010},\ket{1001}\nonumber \\
 & \ket{0110}, \ket{0101},\ket{0011}
\}.
\label{eq:subsec:hardware:bases}
\end{eqnarray}
The number of bases is not further reduced; that is, the effective Hamiltonian constructed here is equal to the full Hamiltonian within the $N_F=2$ electron number. The above set can be interpreted as the collection of all the \textit{S} and \textit{D} exciations, as introduced in sub-section \ref{subsec:fockspace},  above a classical configuration such as $\ket{1100}$.

The Hamiltonian of the problem, stated as a weighted sum of Pauli string operators, as in Eq.~(\ref{eq:subsec:mapping:JW-H}), has a total of fifteen operators, which can be grouped as:
\begin{eqnarray}
G_1 &=& 
\{IIII, \, ZIII, \, IZII, \, IIZI, \, IIIZ, \, 
  ZZII, \nonumber \\
& &
ZIZI, \, ZIIZ, 
  IZZI, 
  IZIZ, \, IIZZ \}
\label{eq:subsec:hardware:group1}
\end{eqnarray}
and 
\begin{eqnarray}
\label{eq:subsec:hardware:group2}
G_2 = \{YXXY, \, XXYY, \, YYXX, \, XYYX\}.
\end{eqnarray}
Here, a Pauli string operator such as 
$ZIII$ is a shorthand for $\hat{Z}_1 \otimes 
\hat{I}_2 \otimes 
\hat{I}_3 \otimes 
\hat{I}_4$, as introduced in Eq. (\ref{eq:subsec:mapping:pauliop}).\par 

Our goal is to use the quantum hardware to evaluate the values for
$\bra{\bf n} \hat{h}_s \ket{\bf n^\prime}$, where $\hat{h}_s$ belongs to $G_{1}$ or $G_{2}$, and $\ket{\bf n}$ and  $\ket{\bf n^\prime}$ can be any of the bases listed in Eq.~(\ref{eq:subsec:hardware:bases}). \par

Note that this task is not a difficult one for a classical computer. However, the objective of the paper and hardware demonstration is to establish a hybrid quantum-classical computational scheme in which the quantum processor performs some of the computational tasks. In the  simulations of this paper, the task is the evaluation of the phase associated with the exchange of fermions; since $\bra{\bf n} \hat{h}_s \ket{\bf n^\prime}$ can be understood as exchanging fermions on $\ket{\bf n}$ via the operator $\hat{h}_s$, and to arrive at $\ket{\bf n^\prime}$. 
The output can only take discrete values of $\pm1$, $\pm i$, and $0$.
\par

In each group $G_1$ and $G_2$, Pauli string operators commute and thus, theoretically, can be measured simultaneously. However, in order to perform simultaneous measurement, one needs to adjust the computational bases, by applying further quantum gates on the target qubits. The appended measurement circuit further increases the circuit depth \cite{gui2020term}. Thus, practically, the simultaneous measurement may not be advantageous, and, in contrast, induces further noise and decoherence on NISQ devices. Nevertheless, for $G_1$, no further  circuit depth is required, as they are all measured in the $z$-basis. \par

We can further reduce the quantum coprocessor computational load by excluding circuits containing terms that are classically efficient to compute. In eqs.~(\ref{eq:subsec:hardware:group1}-\ref{eq:subsec:hardware:group2}), the operators in $G_1$ do  not contribute to off-diagonal matrix element of the Hamiltonian, while operators in $G_2$ contribute \emph{only} to the off-diagonal matrix elements. This information is used to reduce the essential number of circuits to be run on the hardware. \par

The evaluation of  an element $\bra{\bf n} \hat{h}_s \ket{\bf n^\prime}$ is performed by assigning an appropriate circuit, as introduced in sub-section \ref{subsec:idea}, to a quantum coprocessor. In this work we make use of the IBMQ cloud open devices.\par

Once a circuit is loaded onto a quantum coprocessor, each circuit is executed a several times. The number of times a circuit is run is referred to as the number of shots. In this work, each circuit was executed for a total of $8,000$ shots. At each run, the qubits are measured, and the collection of the realized configurations is used to evaluate  value of $\bra{\bf n} \hat{h}_s \ket{\bf n^\prime}$.

The preparation of the circuits is performed using the transpile and assemble routines available in the IBM Qiskit API \cite{Qiskit}, which enables collating individual circuits, corresponding to the diagonal and off-diagonal terms, to prepare each circuit to run as a single job on a target IBMQ device. In this work, we make use of the IBMQ 5-qubit Valencia hardware \cite{IBMQ-Valencia}. The number of circuits --- for both real and imaginary parts separately ---  after the transpile and assemble routines for the diagonal and off-diagonal terms corresponds to 66 and 60 circuits, respectively. The execution timings for the IBMQ QASM simulator range between 0.21-1.5 seconds and for the IBMQ Valencia 5 qubit device 274-298 seconds with 8000 shots per circuit. Due to the limited number of qubits available on this processor, direct measurement is used, that is by measuring all the qubits at the end of the circuit.
\par

\begin{figure}[ht]
    \centering
    \includegraphics[width=0.45\textwidth]{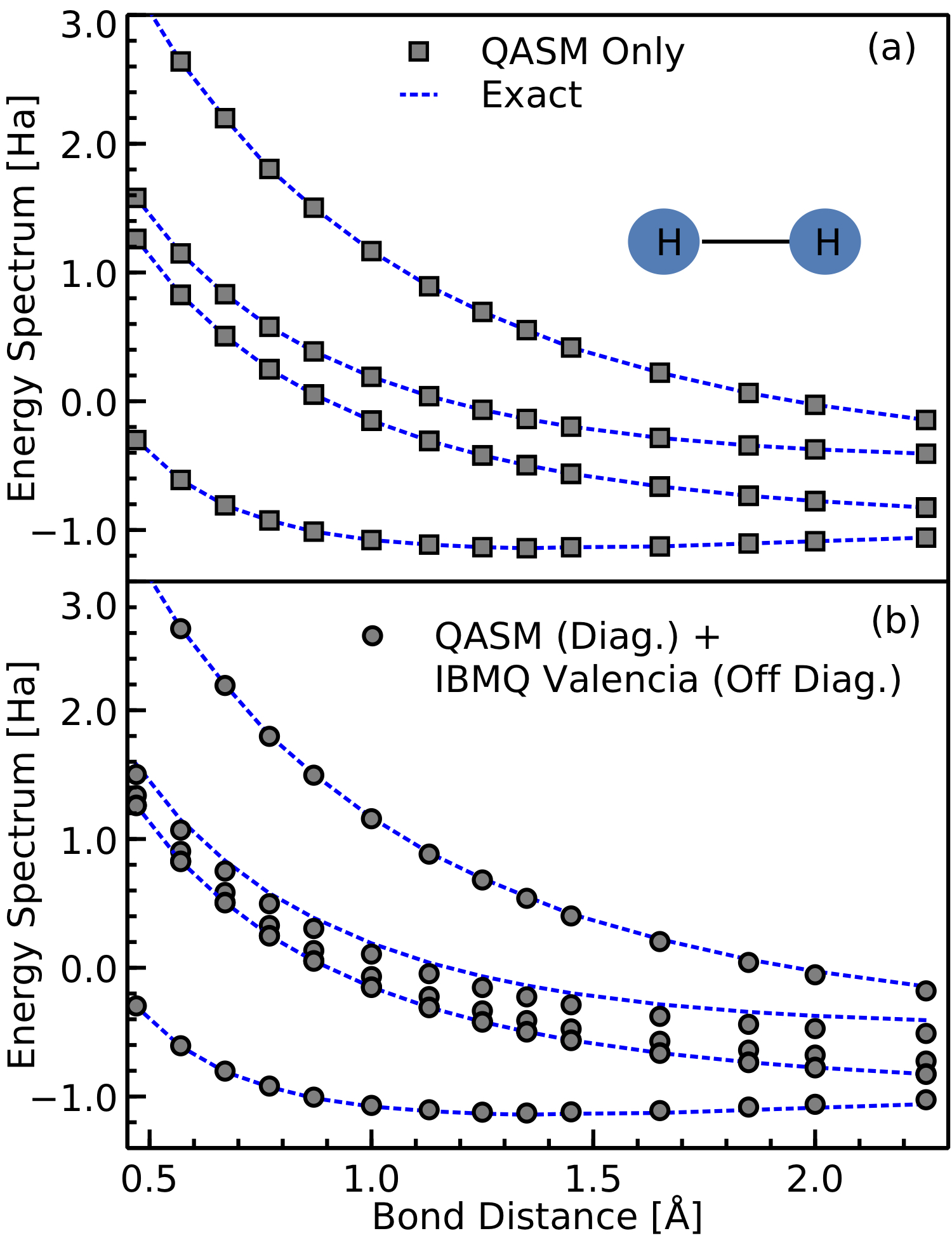}
    \caption{The binding curves for H$_2$ using a (a) noiseless QASM simulator to obtain all Pauli terms and (b) a noiseless QASM simulator to evaluate diagonal Pauli terms and the IBMQ Valencia 5-qubit device to measure off-diagonal terms. 
    }
    \label{fig:H2_QASMHardware}
\end{figure}

As a sanity check, we first use IBMQ QASM simulator \cite{IBMQ-QASM} to perform all steps involved as outlined in \ref{sec:method}. The calculations are shown in Fig.~\ref{fig:H2_QASMHardware}(a). The results perfectly match the exact values as expected. Furthermore, in Fig.~\ref{fig:H2_QASMHardware}(b) the off-diagonal terms in the Hamiltonian are evaluated using IBMQ Valencia device, while the diagonal elements are evaluated using the IBMQ QASM simulator.  In both cases, upon diagonalization of the obtained Hamiltonian, the energy spectrum is in good agreement with the exact values. We observe in Fig.~\ref{fig:H2_QASMHardware}(b) that there is a slight lifting of the double degeneracy of H$_2$, which we associate to the device noise.  This is due to the occurance of non-zero values in the measured off-diagonal terms  when in fact the exact value should be zero.\par

\begin{figure}[ht]
    \centering
    \includegraphics[width=0.45\textwidth]{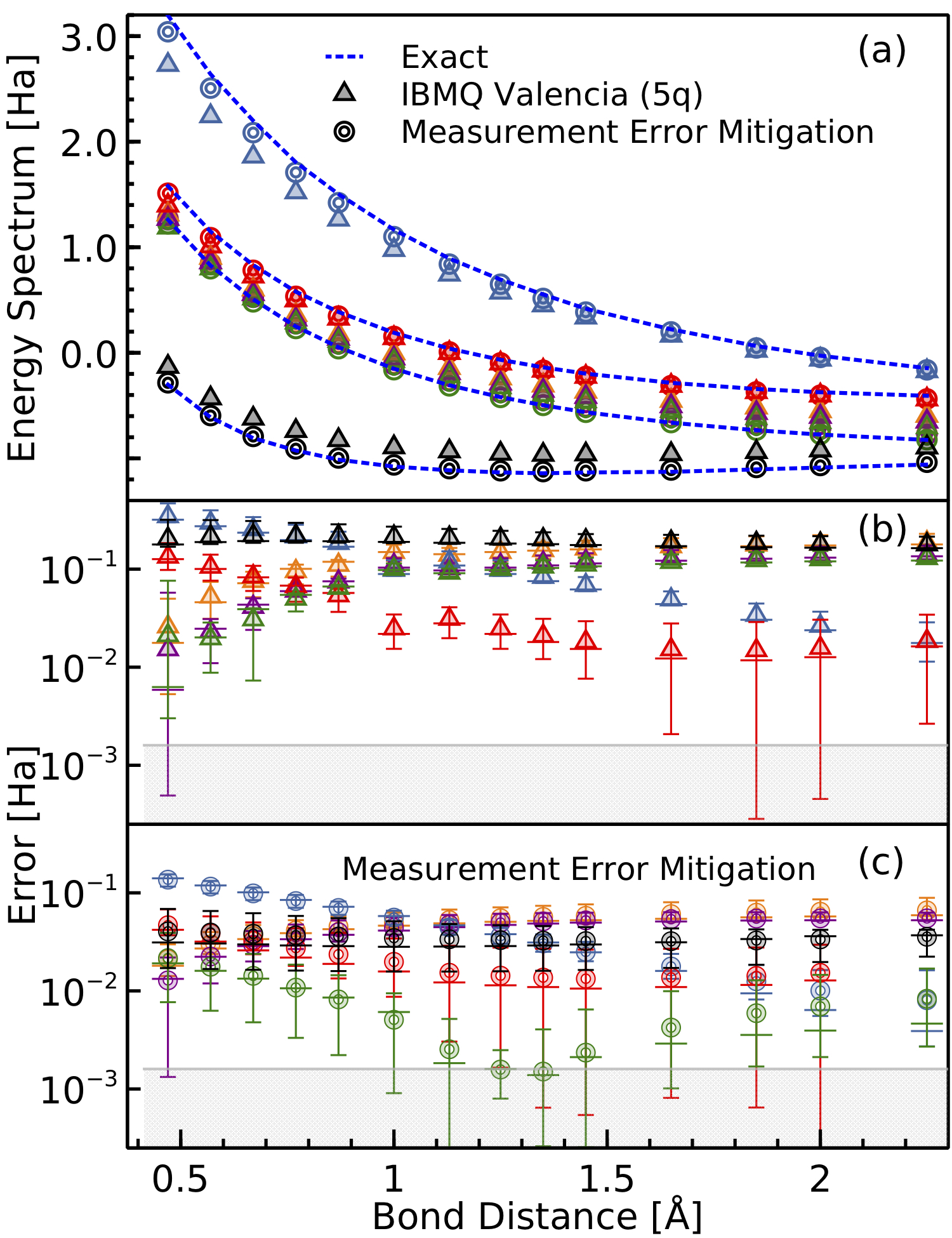}
    \caption{The complete hardware calculation using the IBMQ Valencia 5-qubit device. The lowest value shown in (a) for 5 distinct hardware runs where 8000 shots is used. The absolute error between exact and hardware results is shown in (b), where symbol indicates the mean and  error-bars min/max amongst the 5 runs. (c) the
    same as (b) but with measurement error mitigation for the target hardware. The semi-opaque region in (b-c) indicates chemical accuracy limit.}
    \label{fig:H2_Hardware}
\end{figure}

Finally, all the matrix elements, that is, diagonal and off-diagonal elements, are evaluated using IBMQ Valencia device. The results are shown in Fig.~\ref{fig:H2_Hardware}-(a), which was done with and without measurement error mitigation. The use of measurement error mitigation marginally improves the results. The discrepancy between the exact and hardware values, without and with measurement error mitigation, are shown in \ref{fig:H2_Hardware}(b-c). \par

The effect of noise is ubiquitous in current quantum hardware and is inherently complex and difficult to characterize. However, the result of noise due to the measurement process can be mitigated to some degree by using a calibration matrix  \cite{Qiskit}. The resulting calibration matrix is then used to reduce errors of subsequent circuit measurements. The application of measurement error mitigation is demonstrated in Fig.~\ref{fig:H2_Hardware}(a) with beneficial impact on error shown in Fig.~\ref{fig:H2_Hardware}(c). \par

The calculations in Fig.~\ref{fig:H2_Hardware}, that is construction of the effective Hamiltonian, $\hat{H}_{eff}$,  are performed through five independent IBMQ job submissions. Each time the energy-surface is slightly different, which can be associated to the inherent device errors and perhaps low number of shots used. The error bars in Fig.~\ref{fig:H2_Hardware}(b-c) indicate the min/max range in errors of the five different calculation. \par

%% file: discussion_summary.tex
\section{Discussion and Summary}
\label{sec:discussion_summary}
Recently an abundance of Variational Quantum Algorithms (VQA) have been introduced to solve electronic quantum many-body problems on NISQ devices \cite{Peruzzo2014,Wecker2015,McClean2016,Yuan2019,Parrish2019,Higgott2019,Jones2019,Nakanishi2019}. These VQA proposals mostly rely on optimization of parametric circuits. In this paper we demonstrate that an alternative approach exists,
specifically for quantum chemistry problems,
which does not require an optimization procedure and parametric ansatz. In this approach, the quantum hardware is used only to prepare a many-body quantum state and to efficiently measure the expectation values of certain target observables with respect to this quantum state. From the output measurements, one can construct what we refer to as an effective Hamiltonian matrix, $\hat{H}_{eff}$. Upon diagonalization of $\hat{H}_{eff}$ using classical eigen-decomposition numerical methods one  obtains the ground state and low-lying excited states of the system. \par

The approach introduced in this paper is particularly suitable for NISQ devices where short quantum circuit depth is essential due to lack of error-correction protocols on these devices. An additional important aspect of this approach is that it provides access to the low-lying energy spectra of the system and not just the ground state in comparison with the original VQE process. \par

In the context of VQE and VQE-type algorithms, several attempts have been made to extend the variational approach to excited states \cite{Tilly2020,Santagati2018,Higgott2019,McArdle2019,Somma2019,Parrish2019}. Quantum subspace expansion \cite{McClean2017,Colless2018} for example constructs a set of non-orthogonal bases out of an optimized ansatz, and performs post-processing to obtain excited states. Deflation techniques as described in ref.~\cite{Higgott2019}, constructs a pseudo-Hamiltonian in which the ground state is excluded and orthogonality is enforced through regularization. Successful examples are introduced for some low-lying excited states of LiH \cite{Jones2019}. In all these previous works, optimization of a parametric ansatz is required therefore necessitates an enormous number of quantum circuit executions and sampling. \par 

The approach introduced in this paper is similar to multistate contracted VQE (MC-VQE) in ref.~\cite{Parrish2019}. The main difference is the application of VQE: MC-VQE is obviously a VQE-type algorithm, our approach is distinct in that it does not require a parametric circuit or variational procedure for optimization. In addition, our method differs since it uses supporting quantum circuit resources, i.e., ancilla qubits, in order to perform interference and measurement that is different from the quantum circuit in ref.~\cite{Diker2016} and its generalization in ref.~\cite{Parrish2019}. \par

An extension of our approach to VQE type algorithm is possible. This can be done by appending a set of  parametric gates that act only on the target qubits to the circuits in Fig.~\ref{fig:subsec:idea:core-circ}. Let us denote this part of the circuit with $U(\theta)$, where $\theta$ stands for a set of parameters. Then, it can be verified that, given $\theta$, the final matrix element obtained from the circuit after measurement (see Eq.~(\ref{eq:subsec:hamiltonian:matrix-elements-1}) for example) becomes 
$\bra{\bf n} U^\dagger(\theta) \hat{H} U(\theta) \ket{\bf n^\prime}$, 
compared to 
$\bra{\bf n} \hat{H} \ket{\bf n^\prime}$.
The appended parametric circuit $U(\theta)$ allows one to 
project the Hamiltonian onto  $\mathcal{S}(\theta)=\{U(\theta)\ket{\bf n}\}$, for a given 
$\theta$. This means the $N_s\times N_s$ effective Hamiltonian is now parametric and depends on the value(s) of $\theta$. The optimal parameter(s) are then obtained by minimizing the ground-state energy of the effective Hamiltonian matrix. The reference state $\{U(\theta)\ket{\bf n}\}$ can  be regarded as the contracted reference states introduced in ref.~\cite{Parrish2019}.

One possible limitation of the circuits shown in Fig.~\ref{fig:subsec:idea:core-circ} is the execution of two-qubit gates corresponding to control operations over $n$-qubits (i.e., series of \textsc{CNOT} gates). For NISQ devices with hardware-restricted qubit connectivity, this may require a  number of \textsc{SWAP} gate operations and therefore can increase the circuit depth and subsequent error rates \cite{Venturelli2018,Nash2020}. In essence, the implementation of the circuits described in this paper will depend on the ability to limit circuit depth and associated error rates by NISQ hardware circuit optimization (i.e., scheduling). However, significant improvements in qubit connectivity of various modalities (e.g., ion traps) \cite{Wright2019,Hazra2020} or optimizing quantum circuits against decoherence \cite{Zhang2019,Holmes2020} may blunt this concern. \par

%% file: acknowledgements.tex
\section{Acknowledgement}
The ideas and methodology discussed in Section~\ref{sec:method} have been supported by General Atomics internal R\&D funds. Their application towards quantum chemistry problems as discussed in Section~\ref{sec:numerical} and Section~\ref{sec:hardware} is based upon work supported by the U.S. Department of Energy, Office of Science, Office of Fusion Energy Sciences, under Award Number DE-SC0020249. We thank Mark Kostuk for his guidance and management as PI under this grant. Additionally, we thank David P. Schissel for feedback and guidance at General Atomics.  \par
Circuit diagrams are rendered using the \LaTeX~Quantikz package \cite{Kay2019}, numerical scripts utilize the Python SciPy package \cite{Scipy}, and 2D plots are generated with the Python matplotlib package \cite{Hunter2007}. We acknowledge use of the IBMQ for this work. The views expressed are those of the authors and do not reflect the official policy or position of IBM or the IBMQ team.

\textbf{Disclaimer}: A portion of this report was prepared as an account of work sponsored by an agency of the United States Government. Neither the United States Government nor any agency thereof, nor any of their employees, makes any warranty, express or implied, or assumes any legal liability or responsibility for the accuracy, completeness, or usefulness of any information, apparatus, product, or process disclosed, or represents that its use would not infringe privately owned rights. Reference herein to any specific commercial product, process, or service by trade name, trademark, manufacturer, or otherwise does not necessarily constitute or imply its endorsement, recommendation, or favoring by the United States Government or any agency thereof. The views and opinions of authors expressed herein do not necessarily state or reflect those of the United States Government or any agency thereof.